\documentclass[fleqn,usenatbib]{mnras}

\usepackage{newtxtext,newtxmath}
\usepackage[T1]{fontenc}
\usepackage{ae,aecompl}
\usepackage[colorinlistoftodos]{todonotes}
\usepackage{placeins}

\usepackage{graphicx}	% Including figure files
\usepackage{amsmath}	% Advanced maths commands
\title[No memory of warps in the density structure]{No memory of past warps in the vertical density structure of galaxies}

\author[J. Garc\'ia de la Cruz et al.]{
Joaquín García de la Cruz,$^{1,2}$\thanks{E-mail: j.garciadelacruz@2017.ljmu.ac.uk}
Marie Martig,$^{1}$
Ivan Minchev$^{3}$
\\
$^{1}$Astrophysics Research Institute, Liverpool John Moores University, 146 Brownlow Hill, Liverpool L3 5RF, UK\\
$^{2}$Institut de Ciències del Cosmos (ICCUB), Universitat de Barcelona (IEEC-UB), Martí i Franquès 1, 08028 Barcelona, Spain\\
$^{3}$Leibniz-Institut f\"{u}r Astrophysik Potsdam (AIP), An der Sternwarte 16, D-14482, Potsdam, Germany\\
}

\date{Accepted XXX. Received YYY; in original form ZZZ}

\pubyear{2021}

\begin{document}
\label{firstpage}
\pagerange{\pageref{firstpage}--\pageref{lastpage}}
\maketitle

\begin{abstract}
Warps are observed in a large fraction of disc galaxies, and can be due to a large number of different processes. Some of these processes might also cause vertical heating and flaring.
Using a sample of galaxies simulated in their cosmological context, we study the connection between warping and disc heating.
We analyse the vertical stellar density structure within warped stellar discs, and monitor the evolution of the scale-heights of the mono-age populations and the geometrical thin and thick disc during the warp's lifetime. 
We also compare the overall thickness and the vertical velocity dispersion in the disc before and after the warp. 
We find that for warps made of pre-existing stellar particles shifted off-plane, the scale-heights do not change within the disc's warped region: discs tilt rigidly. For warps made of off-plane new stellar material (either born in-situ or accreted), the warped region of the disc is not well described by a double $\mathrm{sech^2}$ density profile. 
Yet, once the warp is gone, the thin and thick disc structure is recovered, with their scale-heights following the same trends as in the region that was never warped. Finally, we find that the overall thickness and vertical velocity dispersion do not increase during a warp,  regardless of the warp's origin. This holds even for warps triggered by interactions with satellites, which cause disc heating but before the warp forms. Our findings suggest that the vertical structure of galaxies does not hold any memory of past warps.
\end{abstract}


%%%%%%%%%%%%%%%%%%%%%%%%%%%%%%%%%%%%%%%%%%%%%%%%%%

%%%%%%%%%%%%%%%%% BODY OF PAPER %%%%%%%%%%%%%%%%%%

\section{Introduction}

Warps in stellar disc of galaxies were first discovered by \cite{vanderKruit1981SurfaceDisks.}. Large surveys of galaxies have since then shown that stellar warps are present in 40 to 75\% of galaxies, depending on the nature of the sample \citep{Reshetnikov1998StatisticsDisks,Ann2006WarpedGalaxies}. \cite{Schwarzkopf2001PropertiesPerturbations} even found that 93\% of interacting/merging galaxies present warps (vs. 45\% of the non-interacting galaxies). 
Warps are also commonly observed in the HI disc of galaxies \citep{Sancisi1976WarpedGalaxies,Bosma198121-cmTypes.,Heald2011TheObservations}. 
Stellar warps reach amplitudes up to $25^{\circ}$ \citep{Sanchez-Saavedra2003AHemisphere}, and gas warps amplitudes up to $33^{\circ}$ \citep{Garcia-Ruiz2002NeutralWarps}. The duration of warps is difficult to establish from observations, but simulations have found that warps can last from several Myr \citep{Semczuk2020TidallyIllustrisTNG} to several Gyr \citep{Shen2006GalacticInfall}, depending on their formation mechanism. Different formation processes can also explain many of the warps' morphological properties \citep{Saha2006OnGalaxies}, for instance, whether warps are L-shaped (one sided), S-shaped (one side of the disc rises, the other declines), or U-shaped (both sides of the disc rise/decline) \citep{Kim2014FormationDisks}. 

In the Milky Way (MW), an HI warp has been detected at galactocentric radii larger than R$\sim$10 kpc, reaching a height $\sim$4 kpc above the midplane in the North (l $\mathrm{\sim90^{\circ}}$) and curving to the South (l $\mathrm{\sim270^{\circ}}$) below 1 kpc \citep{Levine2006TheDisk}. 
A warp in the stellar Galactic disc has also been detected at galactocentric radii larger than R$\sim$9 kpc, reaching heights above the mid plane up to $\sim$5 kpc \citep[e.g.,][]{Lopez-Corredoira2014FlareData,Liu2016TheDisk,Poggio2018TheKinematics,Romero-Gomez2019AstronomyWarp,Cheng2020ExploringDisk,Antoja2021GaiaAnticentre}. 
Simulations suggest that the warp in the Galactic disc is probably due to the gravitational influence of the infall of the Large Magellanic Cloud \citep{Laporte2018ResponseScenario,Laporte2019FootprintsSet,Poggio2020MeasuringSatellite}.

While \cite{VanDerKruit2007TruncationsGalaxies}, using the SDSS survey, suggested that  stellar and gaseous warps are different components, with different evolution scenarios, \cite{Gomez2017WarpsSimulations} found in simulations that gas and stars in warps follow the same pattern, and that they remain coincident for at least 1 Gyr. This seems to be in agreement with what \cite{Chen2019AnCepheids} found in the MW, where the morphology of the warp for the gas and the youngest stellar populations trace each other up to at least 20 kpc, suggesting that the two components are closely linked to each other.

\begin{figure*}
\centering
\includegraphics[width=\textwidth]{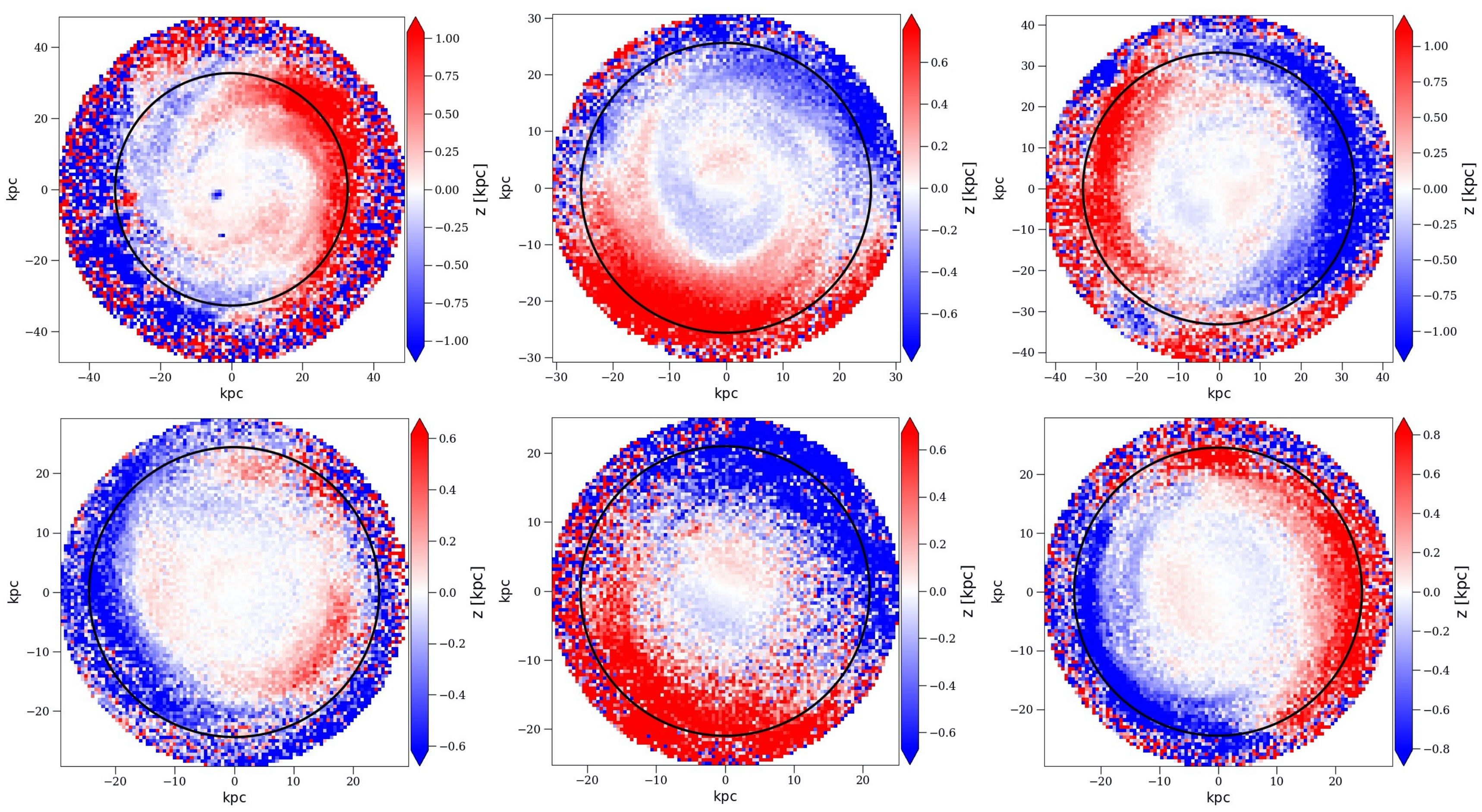}
\caption{A sample of vertical position maps where warps are detected. From left to right and top to bottom: \textbf{g38}, \textbf{g83}, \textbf{g39}, \textbf{g62}, \textbf{g126}, and \textbf{g48}. Each pixel is colour-coded by the stellar particles' mean vertical position of stellar particles. The black circle corresponds to $R_{\mathrm{25}}$, which we use to define the edge of the disc. The left column shows two examples of L-shaped warps (or one-sided), while the rest shows S-shaped warps (or two-sided).}
\centering
\label{fig:warpsample}
\end{figure*}

Many mechanisms have been proposed for warp formation, including misalignment between the disc and the halo \citep{Debattista1999WarpedMomenta,Ideta2000TimeHaloes}, misalignment between the inner stellar disc and the gas disc \citep{Roskar2010MisalignedDiscs,Aumer2013}, interactions between magnetic fields and HI gas \citep{Battaner1990IntergalacticWarps,Battaner1998AWarps}, accretion of intergalactic matter onto the disc \citep{Lopez-Corredoira2002OldWarp,Semczuk2020TidallyIllustrisTNG}, bending modes or waves embedded in the disc \cite[e.g.][]{Sparke1988AWarps,Revaz2004,Chequers2017SpontaneousDiscs}, and the gravitational interaction of an in-falling satellite \citep{Ostriker1989WarpedDiscs,Weinberg1998DynamicsCompanion,Jiang1999WarpsInfall}. In the latter scenario, new infalls can regenerate and maintain the warp for several Gyr \citep{Shen2006GalacticInfall}. 

Many of these mechanisms are often associated with disc heating and flaring \citep[e.g.,][]{Gerssen2012DiscSequence,Minchev2015ONDISKS,Pinna2018RevisitingSimulations}. For instance, mergers have been proposed to create disc heating \citep{Toth1992GalacticOmega,Sellwood1998RESONANTGALAXIES,Kazantzidis2008ColdAccretion,Kazantzidis2009ColdAccretion} and disc flaring \citep[e.g.,][]{Villalobos2008SimulationsDiscs,Bournaud2009,Martig2014b}.  
Another example is misaligned infall of gas, which can also cause disc heating and flaring \citep{Scannapieco2009}, as well as warping.
Therefore, disc heating and warping may happen simultaneously. Indeed, the outskirts of the MW's disc seem to be both flaring \citep[e.g.,][]{Amores2017EvolutionShape,Lopez-Corredoira2018,10.1093/mnras/sty2604} and warping \cite[e.g.,][]{Chen2019AnCepheids,Yu2021TheStars}. \cite{Khachaturyants2021HowDiscs} suggested that settled warp stars may contribute to the flaring of the youngest stellar populations observed at large galactocentric radii in the MW. In some cases, it could even be possible that the warps themselves might directly cause some vertical heating:
bending waves, a phenomenon closely related to warping, have been proposed to heat discs \citep{Khoperskov2010NumericalGalaxies,Griv2011Velocity-anisotropy-drivenDiscs}. 
However, heating and warping do not necessarily always happen together. In the case of satellite-galaxy interactions, whether the disc tilts or gets heated depends on different properties of the satellite, including its orbital properties \citep{Velazquez1999SinkingDiscs}. \cite{Shen2006GalacticInfall} similarly found in simulations that in a satellite-galaxy interaction, the host galaxy's disc tilts remarkably rigidly, so that vertical heating is very limited.

Thus, it is worth exploring the connection between warping and disc heating/flaring. In this paper, we use a sample of simulated galaxies from \cite{Martig2012} to explore how much disc vertical heating happens during warps, and how warps affect the vertical stellar density structure of galactic discs. The paper is divided as follows: in section \ref{sec:methods} we describe the suite of simulations used as well as our methods to characterise the warps and the vertical structure of discs. In section \ref{sec:warpsample} we describe different properties of the warps, including how they form. In section \ref{sec:monitoringdischeating}, we address the connection between warps, disc heating, and disc flaring; and summarise our conclusions in section \ref{sec:conclusions}.

\section{Methods}
\label{sec:methods}

\subsection{Simulations}
\label{sec:simulation}
The simulations used in this work belong to a suite of simulated galaxies in their cosmological context from \cite{Martig2012}. 
The technique employed in these simulations is described in more detail in \cite{Martig2009}, and consists of two different stages.
Firstly, a dark-matter-only $\Lambda$CDM cosmological simulation is performed with $512^3$ dark matter particles. This is done using the adaptive mesh refinement code \texttt{RAMSES} \citep{Teyssier2002}.
In this simulation, we identify the dark matter halos with a final mass between 2.7$\times 10^{11}$ and 2$\times 10^{12}$ $\mathrm{M_{\odot}}$ that live in isolated environments. The merger histories and diffuse dark matter accretion are recorded for those halos from $z=5$ to $z=0$.  
Secondly, a re-simulation follows the growth of a seed galaxy which evolves from $z=5$ to $z=0$ using the merger and accretion histories obtained in the first simulation. A galaxy containing stars, gas, and dark matter replaces each incoming halo. For this simulation stage, a Particle-Mesh code \citep{Bournaud2002,Bournaud2003} is used, where gas dynamics is modelled with a sticky-particle algorithm.
Each simulation box is 800 $\times$ 800 $\times$ 800 $\mathrm{kpc}$ in size, and the mass resolution is 1.5$\times 10^{4}$ $\mathrm{M_{\odot}}$ for gas particles and stellar particles formed during the simulation, 7.5$\times 10^{4}$ $\mathrm{M_{\odot}}$ for stellar particles in the initial seed galaxies at $z=5$, and 3$\times 10^{5}$  $\mathrm{M_{\odot}}$ for dark matter particles. The spatial resolution is 150 pc. Star formation obeys a Schmidt-Kennicutt law with an exponent of 1.5 \citep{KennicuttJr.1998} above gas densities of 0.03 $\mathrm{M_{\odot} pc^{-3}}$. Energy feedback from supernovae is included and a mass loss scheme is implemented following that of \cite{Jungwiert2001} and used in \cite{Martig2010}.

\begin{figure}
\centering
\includegraphics[width=\linewidth]{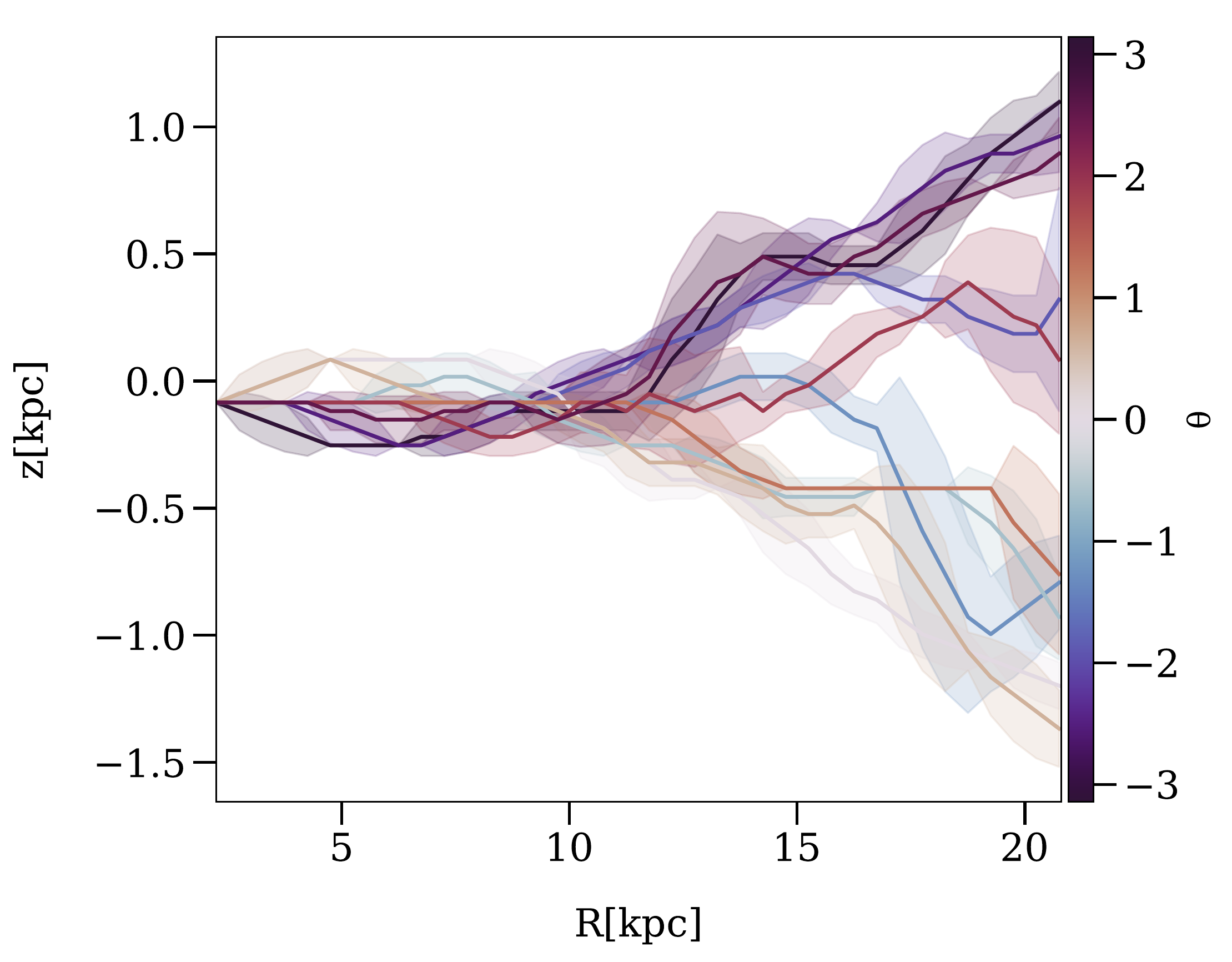}
\caption{Mean vertical position of the stellar density peak as a function of radius colour-coded for different azimuthal angles for galaxy \textbf{g126}. Shaded areas represent the dispersion around the mean vertical position. The values at the largest radius are used to compute the maximum warp height and angle given in Table \ref{tab:example}. We define the onset radius of the warp as the radius where the vertical position of the stars starts deviating from 0 for different azimuthal angles (this would be $\sim$11 kpc for galaxy \textbf{g126} shown here).}
\centering
\label{fig:warp_char}
\end{figure}

\begin{figure}
\centering
\includegraphics[width=\linewidth]{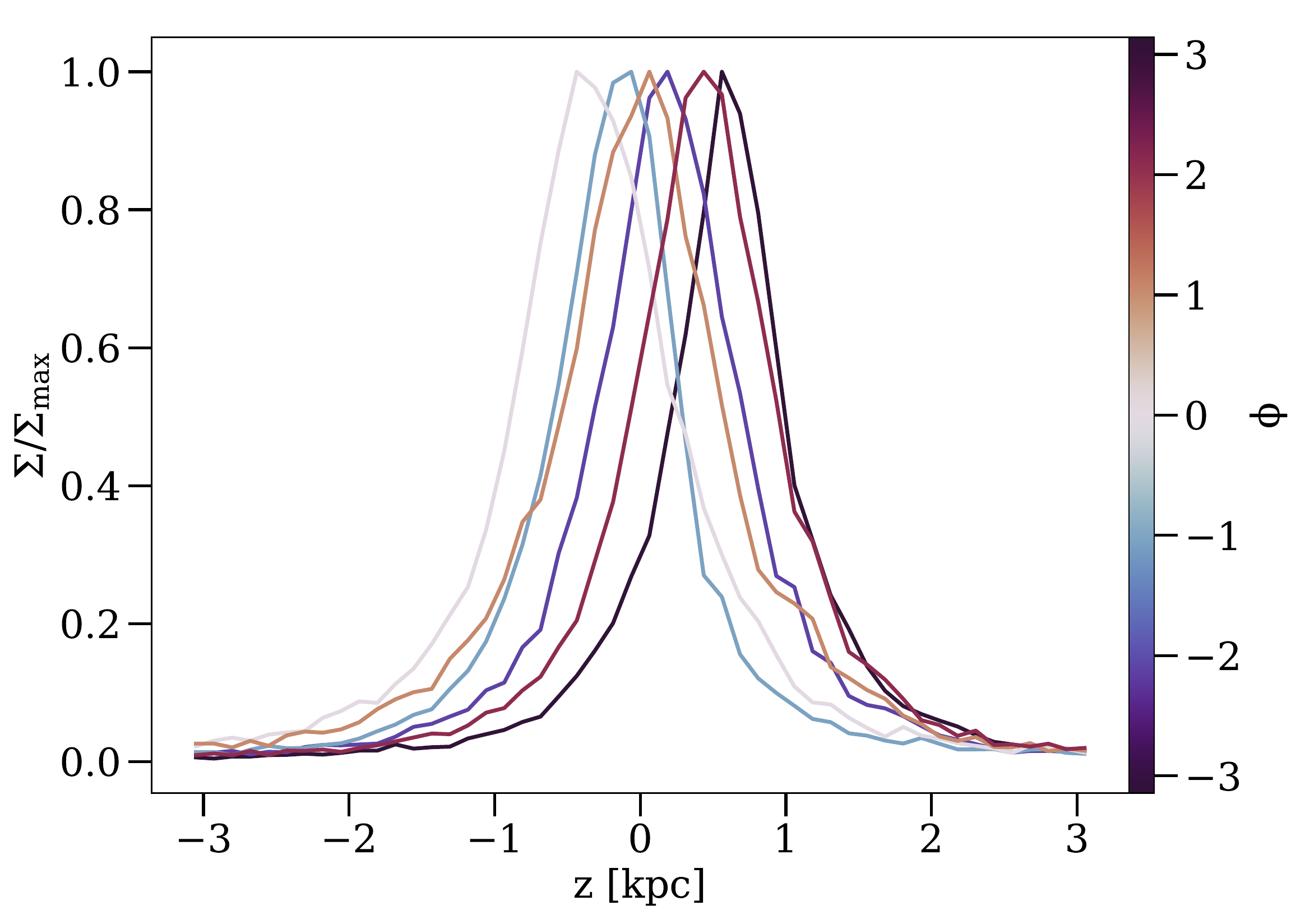}
\includegraphics[width=\linewidth]{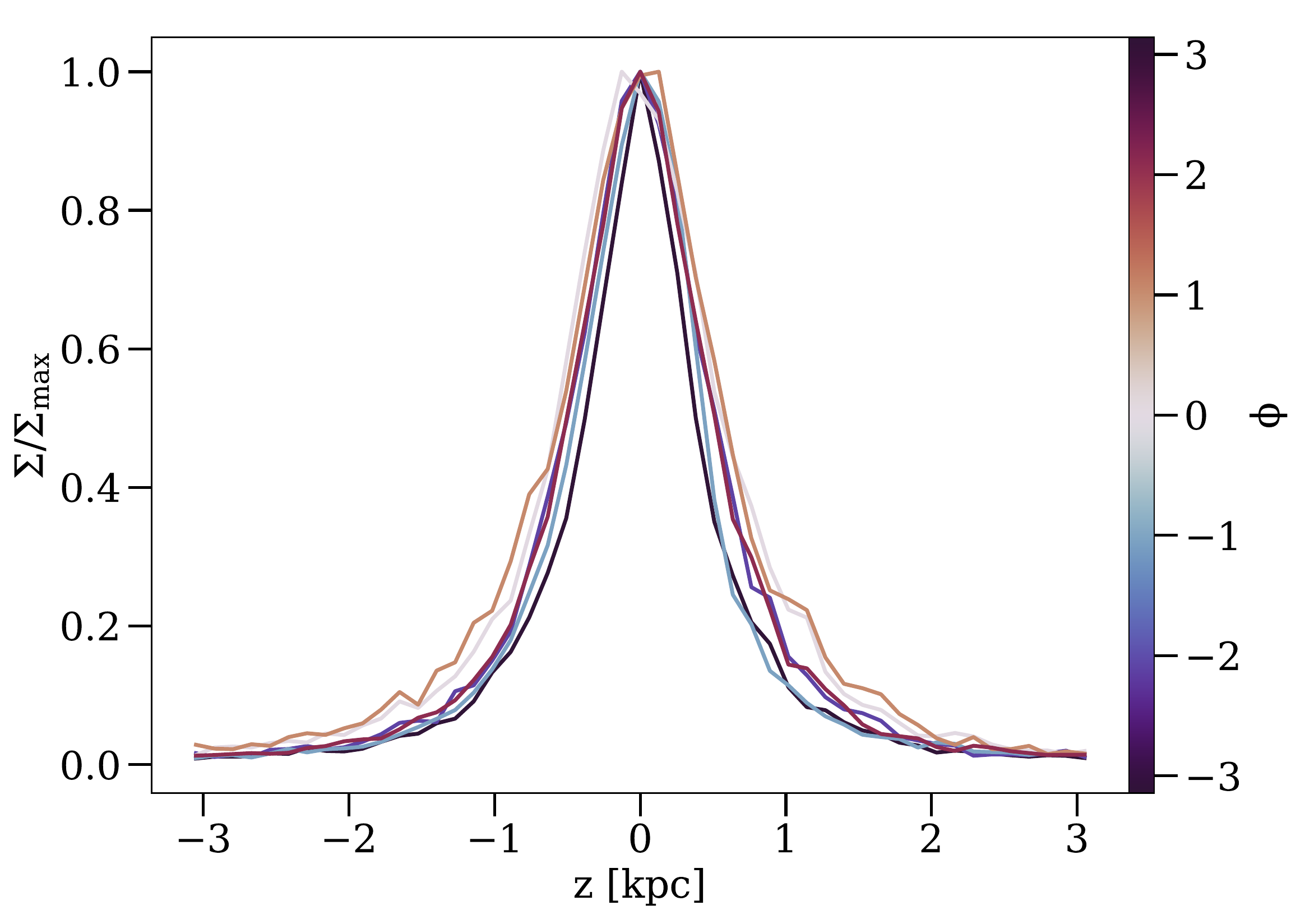}
\caption{\textit{Top:} normalised vertical stellar density for different azimuthal angles within a stellar ring in the disc for galaxy \textbf{g83}. \textit{Bottom:} the result of shifting uniformly all stellar particles in the given azimuthal angle following the method described in Sec. \ref{fig:shiftcorrection}. This shows density profiles that are very similar in the different angular sectors of the disc, with only small variations at large height. We apply this methodology for all the analysis of this work.} 
\centering
\label{fig:shiftcorrection}
\end{figure}

\begin{table*}
\centering
\caption{Summary of the general features of our warp sample. In the final time column, the symbol ``-'' means the warp is ongoing by $z=0$. In the symmetry/shape column, ``L'' means the warp is one-sided, while ``S'' means is two-sided, with each side going in opposite directions.}
\begin{tabular}{lcccccc}
\hline
galaxy ID & initial time & final time & symmetry/shape & max. |z| & max. $\mathrm{|\theta|}$ & likely warp agent\\
& Gyr & Gyr &  & kpc & $\circ$ \\
\hline
g38 & 11.9 & 12.4 & L & 1.40 & 2.45 & internally driven \\
g128 & 9.8 & 10.1 & L & 0.61 & 2.10 & internally driven \\
g35 & 13.1 & - & L & 3.07 & 6.08 & satellite \\
g62 & 13.0 & - & S & 1.59 & 3.74 & satellite \\
g83 & 12.3 & - & S & 1.69 & 3.76 & satellite \\
g126 & 12.1 & - & S & 1.97 & 6.12 & satellite\\
g148 & 11.9 & 13.3 & S & 0.63 & 4.08 & satellite \\
g48 & 9.8 & - & S & 5.43 & 12.25 & stellar ring \\
g59 & 11.1 & - & S & 7.60 & 21.68 & stellar ring \\
g102 & 9.8 & 12.6 & S & 2.93 & 8.9 & stellar ring \\
g39 & 8.0 & 9.7 & S & 6.69 & 11.55 & accretion \\
\hline
\end{tabular}
\label{tab:example}
\end{table*}

\subsection{Warp selection and characterization}

\subsubsection*{Alignment of the galactic plane}
\label{sec:warpselection}
In order to study warps, we need to globally align the galactic disc in the XY plane. For this, we first compute the total angular momentum of all stellar particles inside a sphere of radius 20 kpc. We then rotate all particles to align this angular momentum with the Z axis. To refine the alignment of the disc in the XY plane, we rotate the disc a second time by computing again the total angular momentum's direction but, this time, within a cylinder with radius equal to $R_{\mathrm{25}}$ and height equal to 6 times the estimated disc thickness $h_\mathrm{scale}$. This estimated disc thickness is defined as the standard deviation of the vertical position of stars located at half the optical radius of the galaxy, $R_{\mathrm{25}}$, also used and described in \cite{GarciadelaCruz2021OnSimulations}. 

\subsubsection*{Identifying warps}

After we ensure the discs are properly aligned with the XY plane, we create maps of the mean vertical position of stellar particles (see examples in Fig. \ref{fig:warpsample}), and use those maps to visually identify the presence of warps. We only select warps that clearly start within $R_{\mathrm{25}}$. We track each galaxy as a function of time, and note when warps start and end. Our final sample consists of 11 warps: 5 that have ended by redshift $z=0 $ and 6 that are still ongoing.

\subsubsection*{Characterizing warps}
\label{sec:warpcharacterization}

Once we have identified the snapshots where the galactic disc is warped, we proceed to characterise the warps. For that, we divide the disc into 10 azimuthal sectors, and each sector into overlapping radial bins 2 kpc wide every 0.5 kpc. For each of these azimuthal and radial sectors, we bin the stellar particles into 100 vertical bins up to 5 times $h_\mathrm{scale}$ above and below the galactic plane, and identify the peak of the vertical density. An example of this can be seen in Fig. \ref{fig:warp_char}. 

We compute the maximum height above the midplane reached by the warp. We also compute the angle between the galactic plane and a line joining the maximum of the warp to the galactic center: we will call this the warp angle.

From all the information extracted in this process and represented in Fig. \ref{fig:warp_char}, we also extract the onset radius of the warp, as it can clearly be seen when the peak of the vertical stellar density starts to shift away from the galactic plane. We record the value of the warp angle, height and onset radius throughout the warp's life.

\subsection{Vertical density profiles \& scale-heights}
\label{sec:scaleheight_fits}

We analyse the vertical density distribution for stars in the disc both inside and outside the warped region of the disc. To compute a single vertical density profile at a given radius, we first need to correct for the different vertical shifts in vertical density as a function of azimuth caused by the warp itself. For that, after dividing the disc into different azimuthal and radial bins as described in Sec. \ref{sec:warpcharacterization}, we compute the dispersion of the vertical position of the stars in every disc region. Then, we compute the median value of the vertical position from all stellar particles whose vertical position is smaller than the previously computed vertical dispersion. Finally, we shift uniformly all the stellar particles within the region by this median value. With this, the peak of the vertical stellar density is aligned with the galactic plane of the galaxy for a given radius and azimuthal sector. An example of this technique is shown in Fig. \ref{fig:shiftcorrection}. 

After correcting for the vertical shift caused by the warp in different azimuthal sectors, we fit the density profiles with a double $\mathrm{sech^2}$ representing a thin and thick disc following the same method as in \cite{GarciadelaCruz2021OnSimulations} and explained below.
We bin the disc stellar particles in cylindrical shells with a width of 2 kpc  and a height of 6 $h_\mathrm{scale}$. 
At each radius, we compute the vertical number density of particles using 30 bins, and fit the profile using a combination of two $\mathrm{sech^2}$ functions:

\begin{equation}
    N(z)=N_{0} \left((1-\alpha)\textrm{sech}^2\left(\frac{z}{h_\mathrm{thin}}\right)+\alpha  \textrm{sech}^2\left(\frac{z}{h_\mathrm{Thick}}\right)\right)
\end{equation}

where $N_{\mathrm{0}}$ is the stellar number density at the mid-plane, $h_{\mathrm{thin}}$ and $h_{\mathrm{Thick}}$ correspond to the scale-heights of the thin and thick disc respectively, and $\mathrm{\alpha}$ is the number density fraction of the thick disc over the global disc. 
Following the work of \cite{Bennett2019VerticalDR2}, we use a Poisson distribution for the likelihood, which we write as follows:

\begin{equation}
    \textrm{ln}\mathcal{L}(N_{c}|N_{p})=\sum-N_{p}+N_{c}\cdot \textrm{ln}(N_p) - \textrm{ln}(N_{c}!) 
\end{equation}

where ${N_\mathrm{c}}$ is the number of stellar particles counted in the bin, $N_{\mathrm{p}}$ is the number of stellar particles predicted by the model, and $N_{\mathrm{c}}!$ is independent of the models and thus ignored for the computation of the likelihood.

The fits for the scale-height are obtained using the Markov Chain Monte Carlo (MCMC) \texttt{python} package \texttt{emcee} \citep{Foreman-Mackey2013EmceeHammer}, with 200 walkers and 5000 steps. The walkers start from random positions around the best fit value obtained using the  \texttt{ScyPy} routine \texttt{curve\_fit} \citep{Virtanen2019SciPyPython}.
As for the priors, we let $h_{\mathrm{thin}}$ and $h_{\mathrm{Thick}}$ take any value from 0 to 15 kpc, $\alpha$ from 0 to 1, and we set $N_{0}$ to be positive. 
The final values we report for each parameter are the median and the 16$^\mathrm{th}$ to 84$^\mathrm{th}$ percentiles range of the posterior distribution. 

We also compute the scale-heights of mono-age populations by fitting a single $\mathrm{sech^2}$ as discussed in \cite{GarciadelaCruz2021OnSimulations}. The mono-age populations are obtained by splitting the stellar particles into 0.5 Gyr age bins from 0 to 9 Gyr, and 2 Gyr age bins from 9 to 13 Gyr. The spatial bins are the same used for the fits of the global thin and thick discs.

\section{Warp formation processes and characteristics}
\label{sec:warpsample}

After identifying and characterising the galactic warps as described in Sec. \ref{sec:methods}, our final sample consists of 11 warps: 5 that have ended by redshift $z=0 $ and 6 that are still ongoing. In this section, we first present the mechanisms creating the warps, and then discuss the shape, duration and amplitude of the warps.

We find that the different warp formation processes can be grouped into two categories. Warps belonging to the first group arise when the stellar particles living in the disc experience a shift of their vertical position. In order words, the warp is made of stellar particles that already existed in the disc prior to the warp itself. The two main channels to create such a warp are or through an interaction with a satellite galaxy, i.e. a flyby or a merger. On the other hand, warps belonging to the second group form when new stellar material is added to the galaxy in off-plane orbital configurations. This can happen either because stars are accreted onto the disc, or because stars are born already in an off-plane configuration out of a tilted gas disc. Below, we provide a more detailed description of the formation of the warps from our sample (see also a summary in Table \ref{tab:example}): \newline

\noindent
\textit{Internally-driven warps}

In two galaxies, we notice warps that appear while no massive satellite is visible in our stellar density maps. A bending wave starts expanding from the inner disc outwards. When this bending wave reaches the outermost part of the disc, it has widened enough to form an L-shaped warp. 
The origin of this feature is difficult to determine: it could be triggered by some of the dark matter halos orbiting the galaxy \citep{Widrow2014BendingDisc,Chequers2018BendingSubstructure}, but could also be driven by spiral arms \citep{Faure2014}, or result from an internal bending instability \citep{Revaz2004}.
We find two warps belonging to this category: \textbf{g38}, \textbf{g128}. \newline

\noindent
\textit{Interactions with satellites}

These warps are created by the interaction between a satellite and the host galaxy, either a merger or a flyby. Galaxies belonging to this category mostly form S-shaped warps. 

\textbf{g35}: a massive satellite crosses the disc at t$\sim$12.8 Gyr and shortly after this, a warp appears at t$\sim$13.1 Gyr on the side of the disc through which the satellite has passed (this is the only L-shaped warp of this category).

\textbf{g62}: a satellite crosses the disc at t$\sim$11.8 Gyr creating multiple vertical bending waves. Around t$\sim$13.0 Gyr, a warp develops on one side of the disc (L-shaped) and then shortly evolves into an S-shaped warp. 

\textbf{g83}: a satellite flies by the disc in a polar orbital configuration at t$\sim$11.4 Gyr, creating a warp at t$\sim$12.3 Gyr. A second pass at t$\sim$13.4 Gyr increases the amplitude of the warp.

\textbf{g148}: a flyby at t$\sim$11 Gyr in a low-inclination orbit passing close to the disc creates a warp at t$\sim$11.9 Gyr.

\textbf{g126}: a merger at t$\sim$10.5 Gyr creates vertical oscillations propagating from the inner to the outer disc. The warp forms at t$\sim$12.1 Gyr when these vertical oscillations reach the outer disc. A stellar ring forms later on, at t$\sim$12.5 Gyr, in a slightly tilted configuration, which reinforces the warp and increases its duration. \newline

\noindent
\textit{Accretion on an inclined orbit}

\textbf{g39}: a 1:20 merger at t$\sim$7.5 Gyr scatters around stellar particles both belonging to the satellite and the host galaxy in off-plane configurations. Around half a Gyr later, at t$\sim$8 Gyr, the disc grows and the off-plane material falls back onto the disc, forming an S-shaped warp.  \newline

\noindent
\textit{Formation of an inclined stellar ring} 

These warps form when a massive merger perturbs the gas disc, which ends up misaligned with respect to the stellar disc. Later on, a flyby triggers star formation in the outer regions of the inclined gas disc, forming an inclined stellar ring. The gravitational interaction between the stellar disc and the stellar ring pulls the latter down to the galactic plane. The warp is the product of this mixing between the aligned stellar disc and the off-plane stellar material. As with warps created from interactions with satellites, these warps are also S-shaped.

\textbf{g48}: a 1:4 merger happens at t$\sim$8 Gyr. About 1.5 Gyr later, at t$\sim$9.8 Gyr, a stellar ring forms following a flyby from another satellite.

\textbf{g59}: a 1:4 merger at t$\sim$7.5 Gyr spreads gas around the galaxy. During the next Gyr, flybys from two massive satellites keep the gas tilted until, finally, it condenses and forms stars at t$\sim$11.1 Gyr. A stellar ring forms in a tilted configuration, and as with \textbf{g48}, this stellar ring creates the warp.

\textbf{g102}: a 1:15 merger happens at t$\sim$8 Gyr. About 1 Gyr later, two satellites fly by close to the disc and, shortly after this, an inclined stellar ring develops outside of $\mathrm{R_{25}}$ at t$\sim$9.8 Gyr. This structure progressively falls onto the disc, creating the warp. \newline

We record the shape and duration of all these warps. We also compute the maximum height above the midplane reached by the warp as well as the maximum warp angle. 
All these properties are summarised in Table \ref{tab:example}. 
The values we find for most warp angles are comparable to those found in observations \citep[e.g.,][]{Sanchez-Saavedra2003AHemisphere,Ann2006WarpedGalaxies,Reshetnikov2016GalaxiesWarps} and in other simulations \citep[e.g.,][]{Gomez2017WarpsSimulations}. Only one galaxy, \textbf{g59}, is a clear outlier with a very strong warp when compared with observational studies. Such strong warps are rarely observed, but they exists (see e.g., \citealp{Sanchez-Saavedra2003AHemisphere}), and are not rare in simulations \citep{Kim2014FormationDisks}.

\begin{figure}
\centering
\includegraphics[width=\linewidth]{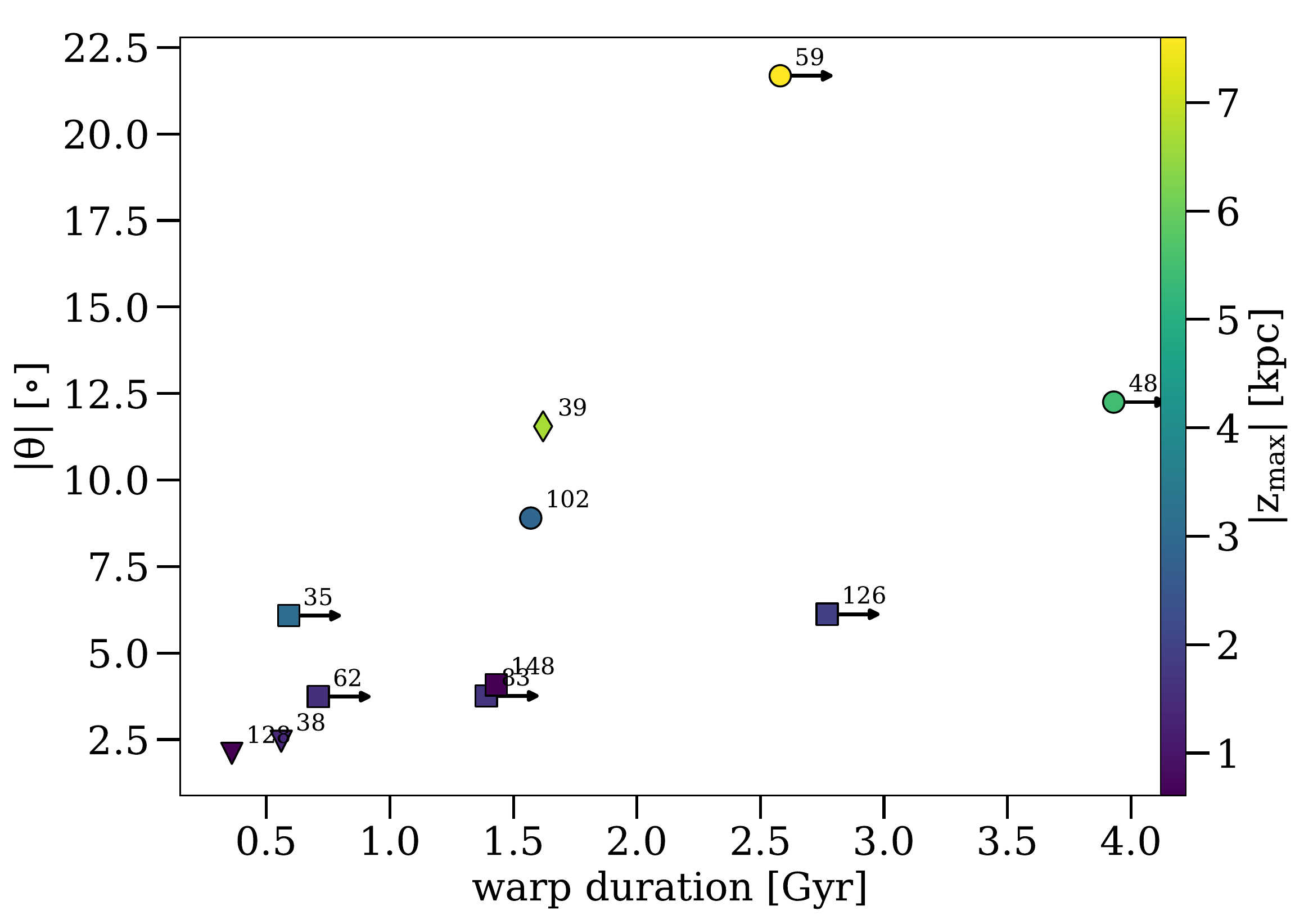}
\caption{Maximum warp angle as a function of the duration of the warp, colour-coded by the maximum height over the galactic plane. The shape of points represents the origin of the warp: triangles are internally-driven, squares are satellite interactions, diamonds are stellar accretion from inclined orbits, and circles are stellar rings being born in off-plane configurations.}
\centering
\label{fig:scatplot_sample}
\end{figure}

In Fig. \ref{fig:scatplot_sample}, we show the warp angle as a function of warp duration (the symbols indicate the origin of the warp, and the colour code the maximum height in kpc reached by the warp). For warps that are ongoing at $z=0$ (indicated by an arrow), the measured duration is only a lower limit, but we still find a general correlations between warp amplitude and duration. We find that internally-driven warps have the shortest duration, around 0.5 Gyr. Their maximum amplitudes are also the smallest within the sample, not reaching above $\mathrm{\sim 2.5^{\circ}}$ or 1.5 kpc. By contrast, warps driven by gravitational interactions between satellites and their host galaxies (flybys or mergers) can last longer than 1 Gyr, in agreement with \cite{Shen2006GalacticInfall}. The longest warp within this category is found in \textbf{g126}: as described earlier, this is a special case where the warp is reinforced by the creation of an inclined stellar ring. Compared to internally-driven warps, warps created by interactions reach slightly higher angles above the mid-plane, up to $\mathrm{\sim 7^{\circ}}$ (corresponding to maximum heights below 3 kpc). This general trend was also observed by \cite{Schwarzkopf2001PropertiesPerturbations}. Finally, warps produced by off-plane stellar rings or by off-plane stellar accretion have the longest duration within the sample, lasting several Gyr. These warps also reach the largest heights above the galactic plane. 

\begin{figure*}
\centering
\includegraphics[scale=0.23]{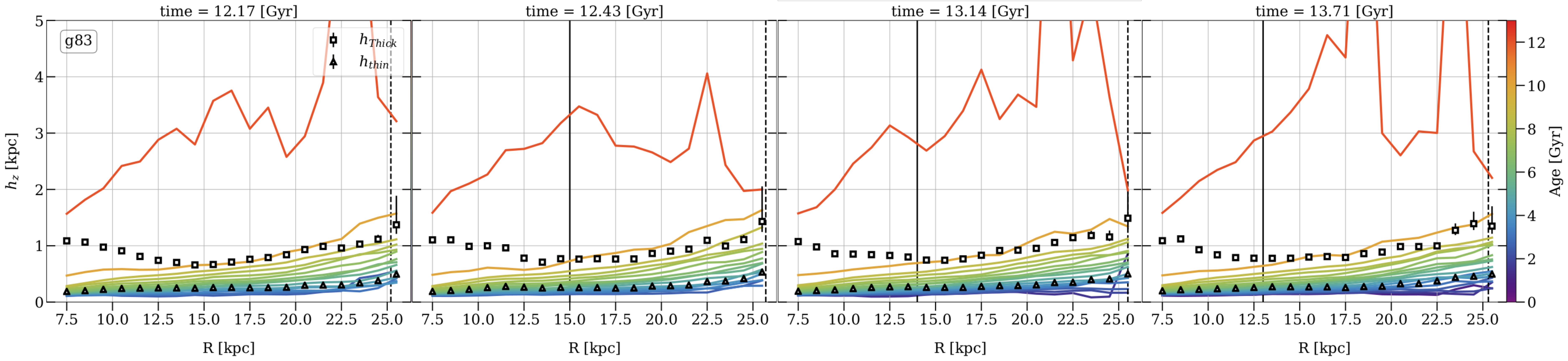}
\includegraphics[scale=0.23]{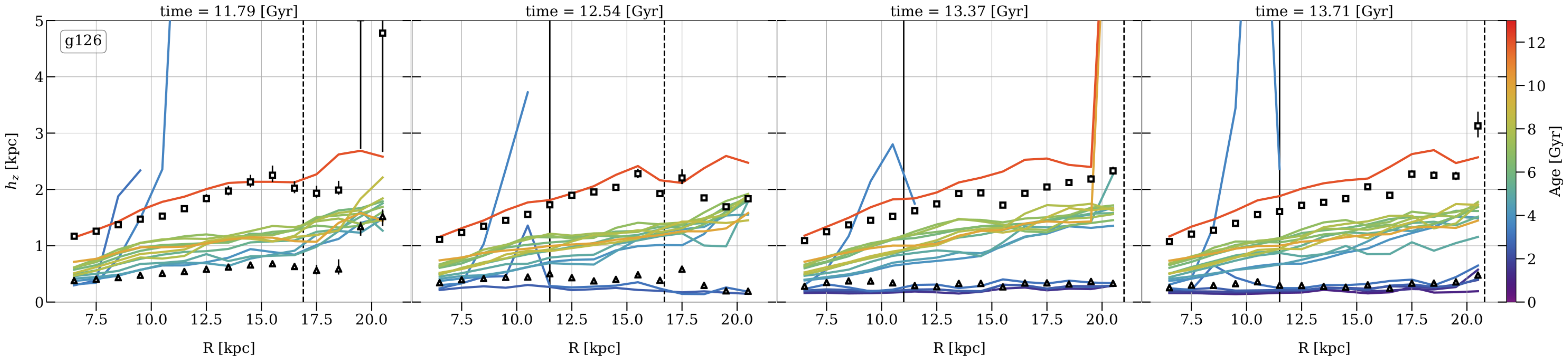}
\includegraphics[scale=0.23]{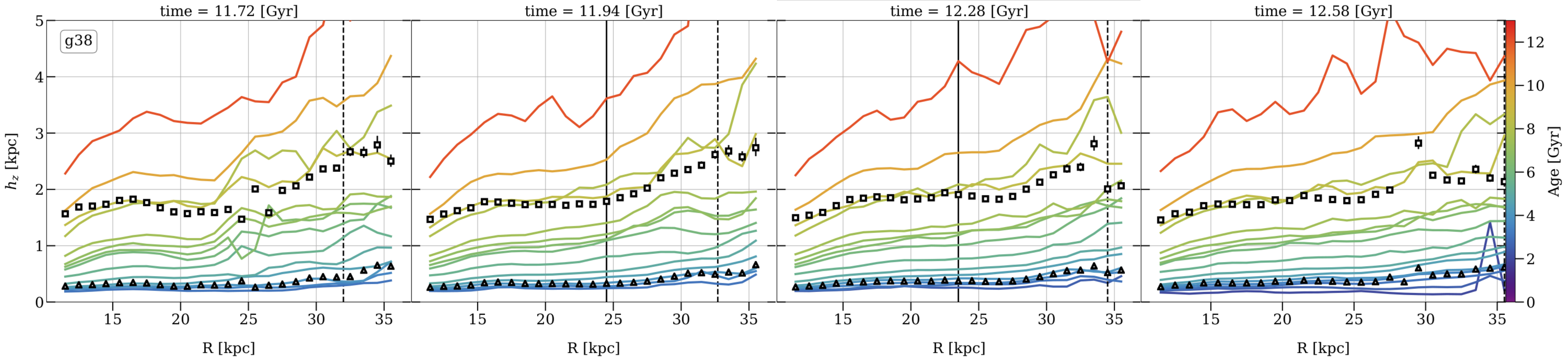}
\caption{Values of the scale-height against radius for the thin disc (triangles), thick disc (squares), and mono-age populations (solid lines) colour-coded by age. Mono-age populations from 9 to 11 Gyr and from 11 to 13 Gyr are binned together, every other mono-age population spans 0.5 Gyr. The x-axis spans from the beginning of the disc to $R_{\mathrm{25}}$ at the cosmological time of the rightmost panel. The solid vertical black line represents the onset radius of the galactic warp, while the dashed vertical black line marks the value of $R_{\mathrm{25}}$ at the time of the snapshot indicated above each panel. The top row represents galaxy \textbf{g83}, where the warps starts at 12.3 Gyr and is still in place by $z=0$. The thick disc stays flat all along. The middle row represents galaxy \textbf{g126}, whose warps starts at 12.1 Gyr and is still active by $z=0$. Both the flaring of the thick disc and MAPs are preserved throughout the warp. The bottom row represents galaxy \textbf{g38}, whose warp starts at 11.9 Gyr and fades away at 12.5 Gyr. The scale-heights of both thin and thick disc, as well as MAPs do not experience major changes during the warp.}
\centering
\label{fig:scaleheightsevolution}
\end{figure*}

\begin{figure*}
\centering
\includegraphics[scale=0.23]{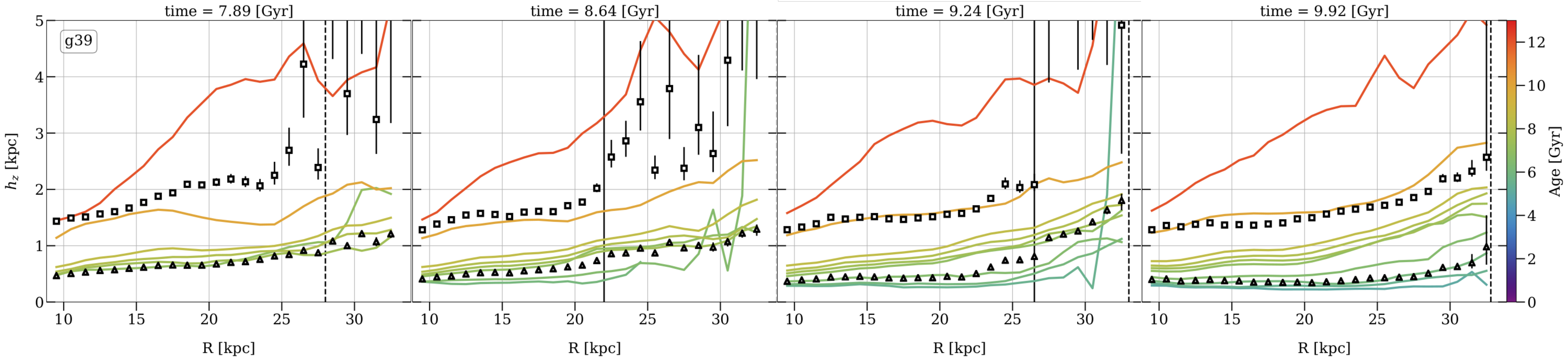}
\includegraphics[scale=0.23]{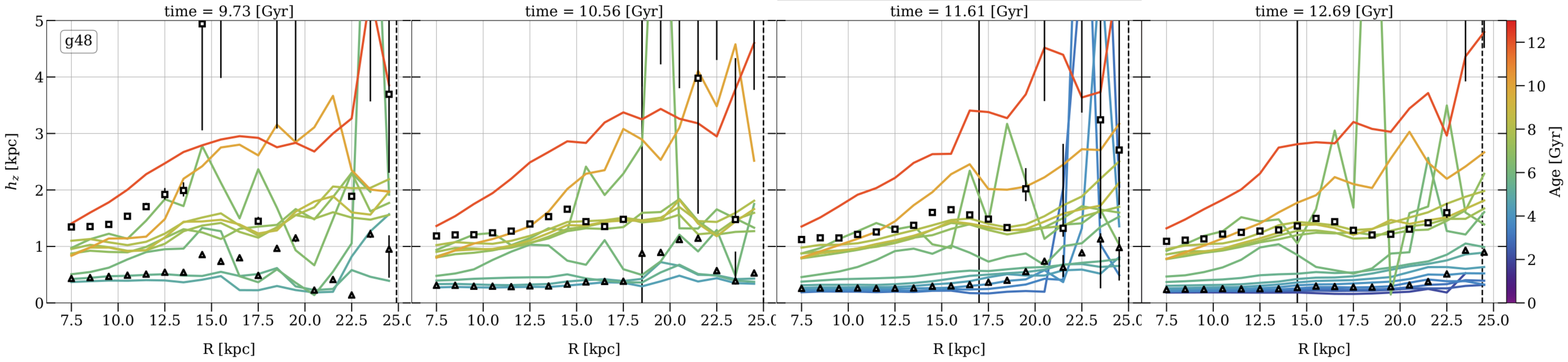}
\caption{Similar to Fig \ref{fig:scaleheightsevolution} but for two galaxies whose disc's vertical density profile is altered during the warps. The top row represents galaxy \textbf{g39}, where the warp starts at 8 Gyr and ends at 9.7 Gyr. In the warped region (beyond the solid vertical line), the vertical density structure is no longer well described by a double $\mathrm{sech^2}$, hence the large uncertainties in the thick disc's scale-heights. However, as the warp fades away, the global thin/thick disc structure is recovered throughout the disc and no sign of the past warp is seen. The bottom row represents galaxy \textbf{g48}, where the warp starts at 9.8 Gyr and is still present at $z=0$. The same effect can be seen, with the thin and thick disc structure recovered for almost all radii even before the warp has finished.}
\centering
\label{fig:exception}
\end{figure*}

\section{Monitoring disc heating during warps}
\label{sec:monitoringdischeating}

In this section, we examine the evolution of the vertical density structure of discs during warps, from 0.5 Gyr before a warp starts to 0.5 Gyr after it ends (for the warps that end before $z=0$). By doing this, we try to minimize the effect that other processes like mergers may have on the disc, and limit our analysis to the warp only. 
We first study whether the vertical density structure of the warped region is distinct from the one in the unwarped inner part of the disc. 
We then follow the time evolution of both the thickness of the disc and the vertical velocity dispersion to understand if the warped region of the disc experiences more vertical heating than the inner disc.

\begin{figure*}
\centering
\includegraphics[scale=0.35]{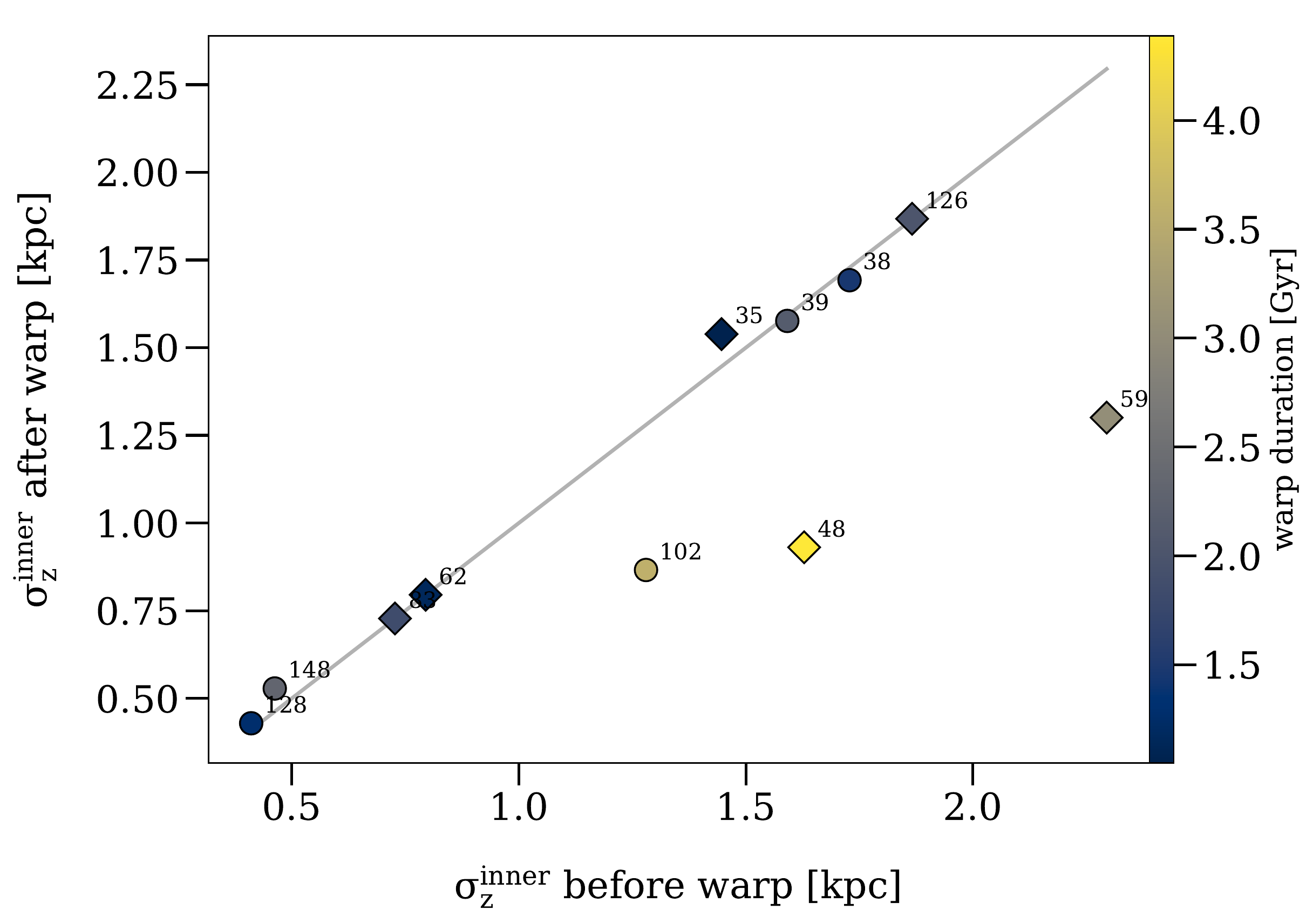}
\includegraphics[scale=0.35]{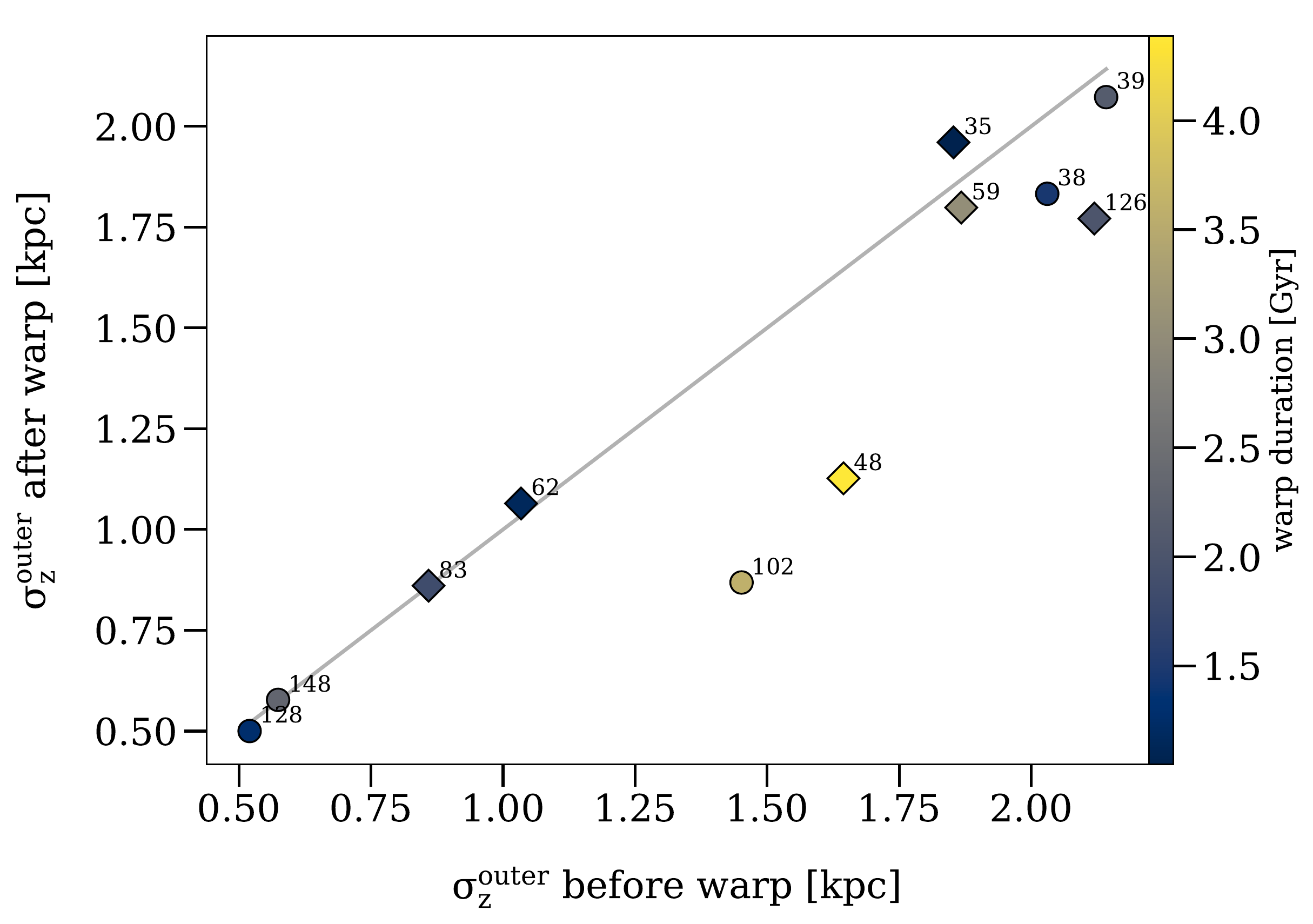}
\includegraphics[scale=0.35]{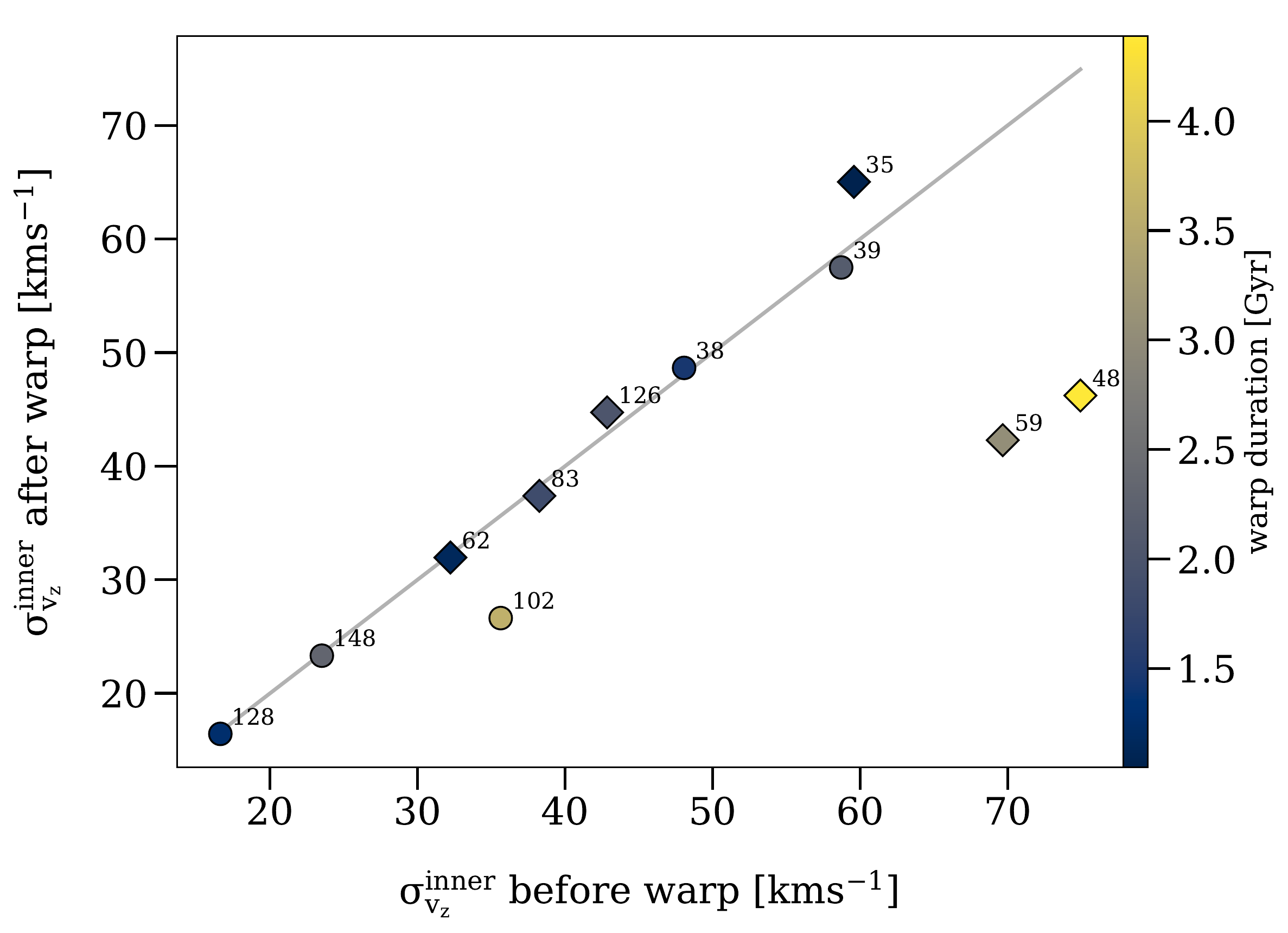}
\includegraphics[scale=0.35]{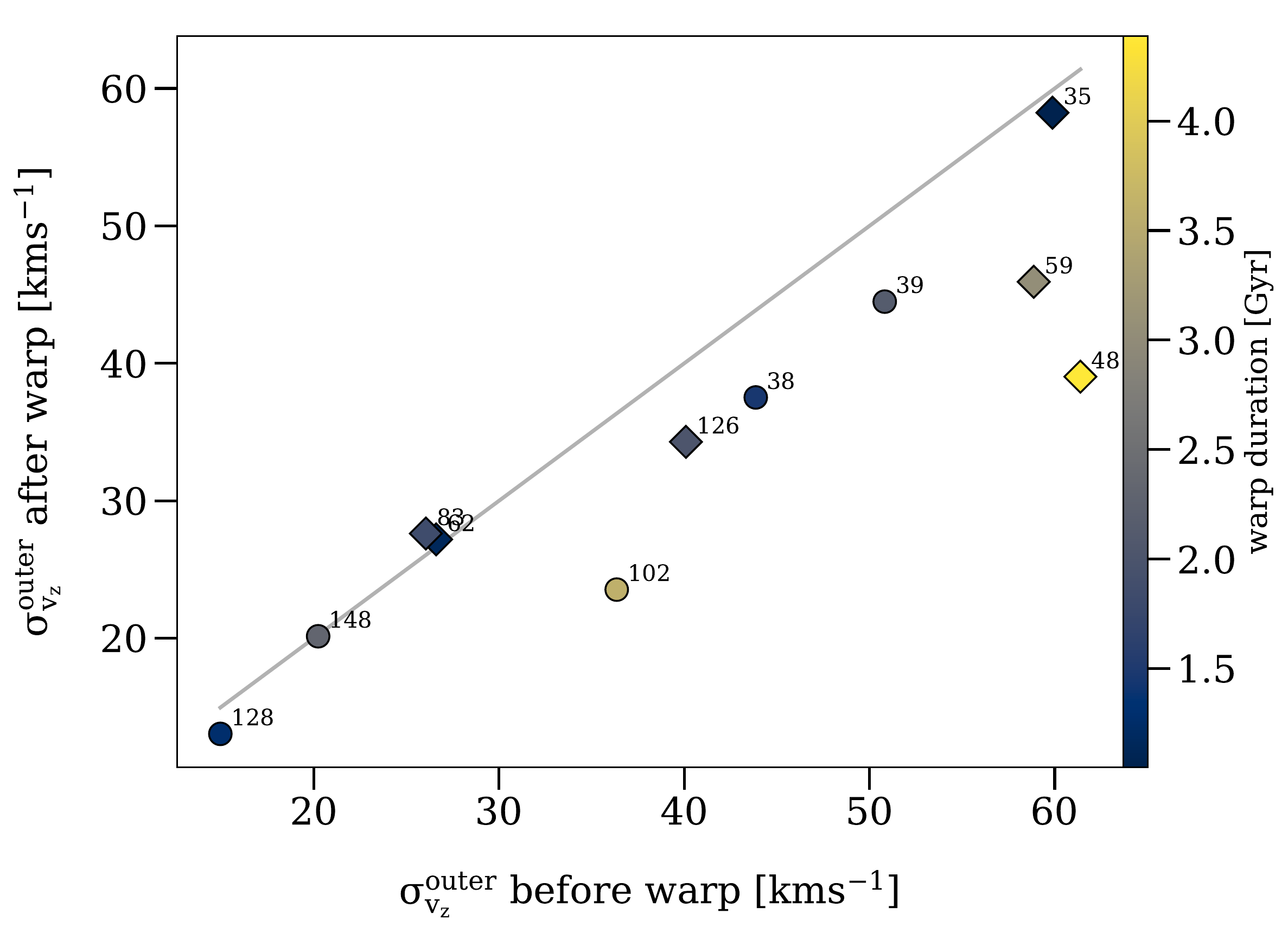}
\caption{\textit{Top row:} standard deviation of the vertical position of the stellar particles before and after the warp within a region in the non-warped part of the disc (left), and outer and warped part of the disc (right). Points are colour-coded by the duration of the warp, and the shape represents whether the warp is finished (circles) or not (diamonds).
\textit{Bottom row}: equivalent of the top row for the vertical velocity dispersion. These plots show that there is no disc heating during the warp. For most galaxies, both the disc thickness and the vertical velocity dispersion are nearly constant as a function of time, while they decrease for warps with the longest duration due to new generations of stars being born in colder configurations during the warp's lifetime.}
\centering
\label{fig:stdz}
\end{figure*}

\subsection{Inner-outer disc transition}
\label{sec:inner_outer}

We examine the evolution of the scale-heights of the thin disc, thick disc, and mono-age populations during different epochs: before the warp starts, during the warp, and (when possible) after the warp has ended. We study whether scale-heights behave differently within the warped region of the disc. We find two main cases within our galaxy sample, which are represented in Figs. \ref{fig:scaleheightsevolution} and \ref{fig:exception} respectively, and explained below: in the first case, the scale-heights never behave very differently within the warped region of the disc. Therefore, in terms of the scale-heights, there is no transition between the non-warped and the warped regions of the disc. In the second case, during the warp, the scale-heights behave very differently within the warped region of the disc. Hence, there is a transition between the non-warped and the warped regions of the disc in terms of the vertical stellar density.

\subsubsection*{The vertical density structure is preserved at all times}
\label{sec:no_transition}

In this category, we find galaxies whose warps are either internally-driven or created by an interaction with a satellite. In those cases, the scale-heights of mono-age populations and of the thin and thick disc behave similarly in the inner non-warped region and in the outer warped region of the disc. For instance, in case of a flat thick disc, \textit{top row} of Fig. \ref{fig:scaleheightsevolution}, the scale-heights of the thick disc are similar in the inner and outer disc. In case of a flared thick disc, \textit{middle row} and \textit{bottom row} of Fig. \ref{fig:scaleheightsevolution}, the scale-heights continuously increase as a function of radius (both for the global thick disc and most of the mono-age populations). The scale-heights sometimes flatten slightly in the very outer disc (outside of $R_{25}$, marked by the dashed black line in Fig. \ref{fig:scaleheightsevolution}), but otherwise they behave similarly in the warped and unwarped regions. For the case of \textbf{g126} in particular, the flaring displayed in the leftmost panel of the \textit{middle row} is the product of the merger at $t\sim10.5$ Gyr. Yet, no increase of flaring is observed during the warp's lifetime. In fact, the rest of the panels of the \textit{middle row} show new generations of stars being born in cold and flat configurations also within the warp.
Lastly, the \textit{bottom row} of Fig. \ref{fig:scaleheightsevolution} is also showing that, after the warp is gone, the vertical structure of the disc does not keep a memory of the warp.

It is interesting to notice that warps in this category are created when stellar particles already living in the disc before the warp get tilted. The fact that the geometrical thin/thick disc structure holds without being strongly affected by the warp means that all the stellar particles are shifted uniformly for a given radius and azimuthal region of the disc. 

\subsubsection*{The vertical stellar density is temporarily altered during the warp}
\label{sec:transition}

In galaxies whose warp is produced by stellar material either being accreted or born in off-plane configurations, the vertical stellar density temporarily changes from the inner and non-warped disc to the outer and warped disc. 
In particular, during the warp, the warped region of the disc is not well described by a double $\mathrm{sech^2}$ density profile, which in practice leads to large uncertainties on the thick disc scale-heights in the warped region (as shown in Fig. \ref{fig:exception}). 

An example of this category is \textbf{g39}, which is illustrated in the \textit{top row} of Fig. \ref{fig:exception}. After a merger at t$\sim$7.5 Gyr, the vertical density within the disc settles down except close to the disc's edge $R_{\mathrm{25}}$ (see leftmost panel). Later, the warp forms when material spread by the merger is accreted.
The amount of material being accreted is such that the stellar vertical density is not well described by a double $\mathrm{sech^2}$ anymore, hence the large uncertainties in the thick disc's scale-heights within the warp (see two middle panels). After the warp is gone, the disc settles back into a double $\mathrm{sech^2}$ structure (see rightmost panel).

A similar inner-outer disc transition is observed for galaxies which formed an outer ring structure off plane, like galaxy \textbf{g48} illustrated in the \textit{bottom row} of Fig. \ref{fig:exception}. At t$\sim$9.7 Gyr, the disc beyond 14 kpc is poorly described by a double $\mathrm{sech^2}$ (see leftmost panel). This is due to the aftermath of a merger at t$\sim$8 Gyr, which induced vertical perturbations (although not in the form of a warp). Yet, once the disc recovers from the merger, the double $\mathrm{sech^2}$ structure starts to reappear. A warp then develops in the  outer disc because of an interaction with another satellite, leading to the formation of an inclined stellar ring. In this region, the disc structure is altered asymmetrically by the high amount of stellar material at high z above or below the galactic plane (see two middle panels). However, once this material settles down within the disc, the double $\mathrm{sech^2}$ reappears even before the warp is fully gone (see rightmost panel).

Regardless of the accreted/in-situ origin of the material, warps that fall into this category are created because of new material is added to the disc at large height, as opposed to those where stars in the disc get tilted.
Here, the double $\mathrm{sech^2}$ structure is not preserved within the warp region, but it reappears after the off-plane material settles down within the disc. Therefore, once the warp is gone, the thin and thick disc structure is recovered, with no sign of where the warp's onset radius used to be.

\subsection{Disc heating}
In order to clarify whether disc heating or flaring take place during warps, we follow the time evolution of the dispersion of the vertical position and vertical velocity of the stellar particles in two radial bins: one in the inner and non-warped part of the disc, and the other in the outer and warped part of the disc. 
We monitor the evolution of these quantities only half a Gyr before and after the warp to minimize the effects of other disc heating agents.  

On the top row of Fig. \ref{fig:stdz} we show the dispersion of the vertical position in a region within the inner disc (\textit{top}) and outer disc (\textit{bottom}). Clearly, the disc thickness does not change as a function of time  for most galaxies, both in the inner and outer region of the discs. The few cases where there is a significant change are the galaxies for which warps last the longest. For these few galaxies, the disc actually becomes thinner with time, because of new generations of stars being born in a much colder configuration. This is seen both in the inner and outer disc.

A similar effect is observed on the bottom row of Fig. \ref{fig:stdz}. The vertical velocity dispersion does not increase because of the warp (both in the inner and outer disc). In many cases, the vertical velocity dispersion does not change at all. For galaxies that we track over longer periods of time, we see $\mathrm{\sigma_{v_z}}$ decrease because of the formation of young stars in a cold configuration.

From this, we conclude that disc heating does not happen during warps, even in the outer regions. This can be due to a few different reasons. First, in cases of warps triggered by interactions, we find that heating happens on faster timescales than warping: warps need a few dynamical timescales to emerge. This means that disc heating is finished before the warp starts. An example of this is galaxy \textbf{g126} in the \textit{middle row} of Fig. \ref{fig:scaleheightsevolution}: a merger happens at t$\sim$10.5 Gyr and creates flaring before the warp starts at t$\sim$12.1 Gyr (see leftmost panel of \textit{middle row} of Fig. \ref{fig:scaleheightsevolution}). Then, the warp develops and no further heating or flaring is observed. Second, not all galaxy-satellite interactions may cause vertical disc heating as pointed out by \cite{Velazquez1999SinkingDiscs}. Finally, while \cite{Khoperskov2010NumericalGalaxies} and \cite{Griv2011Velocity-anisotropy-drivenDiscs} suggested that bending waves can cause vertical disc heating, we find that warps are not a source of vertical heating in our simulations. Thus, warps only correspond to a global shift of the vertical structure of the disc or to an asymmetrical alteration of the vertical density, and when warps disappear discs recover their initial vertical structure.

\section{Conclusions}
\label{sec:conclusions}

In this work, we analysed the connection between disc heating and warping using the simulations from \cite{Martig2012}. We chose a subsample of 6 simulated galaxies with warps present at $z=0$, as well as 5 other galaxies which underwent warps in the past. 
We computed scale-heights for mono-age populations as well as for thin and thick discs after correcting for the warps' vertical shift as explained in Sec. \ref{sec:methods} and represented in Fig. \ref{fig:shiftcorrection}. 

We found that for warps either internally driven or created by interactions, the scale-heights of the geometrical thin and thick disc, and of mono-age populations, are not affected by the warp: whether the thin and thick disc are flat or flared, the trend is always preserved all the way through the disc. In other words, for warps formed when the stars are shifted off-plane, the warp seems to affect uniformly all the stellar disc, and does not alter the structure of the vertical stellar density. We also noted that these warps tend to have shorter lifetimes and reach lower heights above/below the galactic plane.

On the other hand, galaxies whose warps are created by the accretion or the formation of stars in off-plane configurations show a vertical stellar density structure that temporarily changes between the inner and non-warped region of the disc and the warped one. While the non-warped region keeps having a double $\mathrm{sech^2}$ density profile, this is altered asymmetrically in the warped region due to the off plane material. However, once the warp fades away, the double $\mathrm{sech^2}$ is recovered and the trends of the thin/thick disc scale-heights as well as of mono-age populations are coherent throughout the disc, without a signature of a pre-existing warp. Additionally, these warps tend to last longer and reach larger heights above/below the galactic plane than the others.

Finally, we found that for all warps the disc thickness and the vertical velocity dispersion do not increase during the warp, indicating that there is no disc heating happening during warps. This can be either because the disc heating induced by the warp agent takes place before warp formation, or because neither the warp agent nor the warp itself are a source of disc heating.

Our results suggest that depending on their origin, warps may or may not alter the vertical stellar structure of the stellar disc. Nevertheless, once warps disappear, they do not leave any imprint in the stellar populations living in the disc. Therefore, in terms of stellar vertical density structure and vertical heating in the disc, galaxies do not hold any memory of warps they might have experienced in the past. Given the relatively small size of our galaxy sample, further work using larger samples from different simulations would greatly help confirm these results.

\section*{Acknowledgements}

We thank Phil James for his support and for useful discussions.
We would like to acknowledge a LIV.DAT doctoral studentship supported by the STFC under contract [ST/P006752/1].
The LIV.DAT Centre for Doctoral Training (CDT) is hosted by the University of Liverpool
and Liverpool John Moores University / Astrophysics Research Institute.

%%%%%%%%%%%%%%%%%%%%%%%%%%%%%%%%%%%%%%%%%%%%%%%%%%
\section*{Data Availability}
The data underlying this article will be shared on reasonable request to the corresponding author.



\bibliographystyle{mnras}
\bibliography{bib} % if your bibtex file is called example.bib

% \begin{thebibliography}{99}
% \bibitem[\protect\citeauthoryear{Author}{2012}]{Author2012}
% Author A.~N., 2013, Journal of Improbable Astronomy, 1, 1
% \bibitem[\protect\citeauthoryear{Others}{2013}]{Others2013}
% Others S., 2012, Journal of Interesting Stuff, 17, 198
% \end{thebibliography}

%%%%%%%%%%%%%%%%%%%%%%%%%%%%%%%%%%%%%%%%%%%%%%%%%%

%%%%%%%%%%%%%%%%% APPENDICES %%%%%%%%%%%%%%%%%%%%%

%%%%%%%%%%%%%%%%%%%%%%%%%%%%%%%%%%%%%%%%%%%%%%%%%%

% Don't change these lines
\bsp	% typesetting comment
\label{lastpage}
\end{document}